\documentclass[aps,prd,superscriptaddress,nofootinbib,10pt]{revtex4-2}

\usepackage{graphicx}
\usepackage{bm}
\usepackage{mathrsfs}
\usepackage{amsmath,amssymb}
\usepackage{orcidlink}
\usepackage{hyperref}

\newcommand{\nn}{\nonumber}
\allowdisplaybreaks

\begin{document}

\title{Nonlinear response and retarded three-point correlators of charge transport}

\author{Jin Hu\,\orcidlink{0000-0003-1179-4603}}
\email[]{hu-j23@fzu.edu.cn}
\affiliation{Department of Physics, Fuzhou University, Fujian 350116, China}

\begin{abstract}
We study nonlinear response of charge transport in relativistic kinetic theory within the relaxation-time approximation.  For a classical gas of massless charged particles coupled to an external $U(1)$ gauge field, we organize the Boltzmann equation perturbatively in the source and derive a recursive solution for the distribution function to arbitrary response order.  At second order we analytically obtain fully retarded three-point  correlators at generic  frequencies and momenta, verify the Ward identities implied by charge conservation.  We also discuss the general properties of these correlators, including their analytic structure, consistency limits, and relation to Schwinger--Keldysh effective theories of diffusive fluctuations.

\end{abstract}

\maketitle

\section{Introduction}
\label{sec:introduction}

Understanding how a system in thermal equilibrium responds to external perturbations is one of the central problems in statistical mechanics and quantum field theory. In the linear regime, the response is encoded in two-point retarded correlation functions
which determine transport coefficients, the spectrum of collective excitations, and the late-time approach to equilibrium \cite{Kadanoff:1963axw,kapusta2006finite,Kovtun:2012rj}. These correlators are particularly important, as they bridge microscopic dynamics and macroscopic transport phenomena.
A central aspect is that the distribution of singularities of these correlators in the complex frequency plane dictates the real-time dynamics. Poles rooted in hydrodynamics represent long-lived collective modes, such as charge diffusion and sound modes, which govern the late-time relaxation toward equilibrium. In contrast, non-hydrodynamic singularities, whether poles or branch cuts, are tied to fast microscopic degrees of freedom. The characteristic timescale for hydrodynamization and the range of validity of the hydrodynamic description are set by these modes. Hence, to understand how hydrodynamics emerges from an underlying microscopic theory, one must characterize the analytic structure of retarded correlators.


Given the crucial role played by these singularities,  extensive research has been dedicated to uncovering the analytic structure of retarded correlators. For strongly coupled plasmas described by large $N$ thermal $\mathcal{N}=4$ super Yang-Mills theory at large 't Hooft coupling, the correlators are known to possess only poles, i.e., quasinormal modes \cite{Hartnoll:2005ju,Kovtun:2005ev,Grozdanov:2016vgg}. In weakly coupled regimes, however, the non-analytic structures become more diverse. In kinetic theory with the relaxation time approximation (RTA), for instance, the studies have revealed the interplay between hydrodynamic poles and the branch cuts associated with non-hydrodynamic degrees of freedom on the principal Riemann sheet \cite{Romatschke:2015gic,Bajec:2024jez}. Recently, this has also been extended to  the presence of an external magnetic field \cite{Zhao:2026mhx}.
Studies based on momentum-dependent RTA models \cite{Kurkela:2017xis,Brants:2024wrx}, as well as other analytical and numerical analyses \cite{Gavassino:2024rck,Moore:2018mma,Ochsenfeld:2023wxz}, have further demonstrated that branch cuts may become prominent and even gapless, approaching the origin arbitrarily closely. When this happens, non-hydrodynamic excitations can acquire parametrically long lifetimes, and the conventional picture, where non-hydrodynamic modes decay rapidly allowing hydrodynamic modes to dominate and hydrodynamics to emerge, ceases to hold because the two sectors are no longer well separated. Whether such a separation persists is ultimately dictated by microscopic details, in particular the nature of interparticle interactions (hard versus soft scattering), which controls the spectral properties of the linearized collision operator. Consequently, the interplay of non-analytic structures in correlation functions requires a case-by-case discussion \cite{Hu:2024tnn}. 

While linear response and two-point functions have been studied in great detail, nonlinear response and higher-order correlation functions remain much less explored in kinetic theory. Nonlinear effects are essential for understanding non-Gaussian fluctuations, which are sensitive probes of critical phenomena and far-from-equilibrium dynamics \cite{An:2021wof,Chen-Lin:2018kfl,Stephanov:2024xkn,An:2020vri}. Three-point correlation functions, in particular, capture the leading nonlinear response of a system. In the framework of Schwinger-Keldysh \cite{Schwinger:1961,Keldysh:1964,Chou:1985} effective field theory, three-point functions have been computed for diffusive systems by incorporating interactions into the effective Lagrangian \cite{Crossley:2015evo,Glorioso:2017fpd,Abbasi:2024pwz}. Within the bottom-up construction of this approach, the free part of the effective action can be often fixed by the Martin-Siggia-Rose formalism \cite{Martin:1973zz}, while the fluctuation part is constrained by the dynamical KMS symmetry \cite{Crossley:2015evo,Glorioso:2017fpd}.  A top-down construction is also possible \cite{Crossley:2015evo,Glorioso:2017fpd,Haehl:2016pec,Haehl:2016uah,Haehl:2018lcu}. These studies have revealed novel analytical structures, such as branch-point singularities generated by loop effects \cite{Abbasi:2024pwz}, and have established deep connections between response and correlation functions through fluctuation-dissipation relations \cite{Wang:1998wg}. Most of the studies discussed above are formulated for diffusive or hydrodynamic systems and are generally applicable in the hydrodynamic regime of low frequencies and long wavelengths. In particular, \cite{Abbasi:2024pwz} recently resummed the diffusion equation, thereby making it possible to discuss problems beyond the hydrodynamic regime. However, direct calculations of three-point retarded correlators in kinetic theory, starting from the underlying transport equation, remain largely unexplored \cite{Delacretaz:2022ocm}.

In this paper, we fill this gap by studying nonlinear response and three-point retarded correlation functions of charge transport within the kinetic framework. The paper is organized as follows. In Sec.~\ref{sec:kinetic}, we review the kinetic theory in the relaxation time approximation. In Sec.~\ref{sec:linear}, we discuss the linear response and two-point retarded correlators. In Sec.~\ref{sec:nonlinear}, we derive the nonlinear response and compute the three-point retarded correlators.  Their analytic structure is discussed in  Sec.~\ref{sec:analytic}, followed
by consistency limits and physical interpretation in Sec.~\ref{sec:limits}. In Sec.~\ref{sec:comparison}, we discuss the physical implications and compare with the effective field theory approach. Finally, we summarize and outlook in Sec.~\ref{sec:summary}. Some technical details are collected in the appendices.


\vspace{0.3cm}
\noindent\emph{Conventions\,}--- We use natural units $\hbar=c=k_B=1$ and the metric $g^{\mu\nu}=\mathrm{diag}(-1,1,1,1)$.  Greek indices run over spacetime and Latin indices over spatial directions.  The charge of each particle is set to unity.  Fourier transforms are defined with the phase $e^{i(\omega t-\mathbf{k}\cdot\mathbf{x})}$.

\section{Kinetic setup and source expansion}
\label{sec:kinetic}

We consider a dilute classical gas of massless charged particles, $p^0=p\equiv|\mathbf p|$ and $|\mathbf v|=1$, with $\mathbf v=\mathbf p/p^0$.  The on-shell one-particle distribution $f(x,\mathbf p)$ obeys the relativistic Boltzmann equation. The Boltzmann equation is a complex nonlinear integro-differential equation, and exact analytical solutions can be obtained only in very few cases \cite{Bazow:2015dha,Hu:2024utr,Wang:2025wyh,Wei:2026qrk}.  To model the collision kernel while preserving analytical tractability,
  we approximate the collision term by the Anderson--Witting relaxation-time model with a constant relaxation time $\tau_R$
  \cite{Anderson:1974nyl}. In the presence of an external $U(1)$ gauge field, the transport equation takes the following form
\begin{equation}
p^\mu\partial_\mu f+F^{i\beta}p_\beta\frac{\partial f}{\partial p^i}=\frac{p_\mu u^\mu}{\tau_R}\left(f-f_{\mathrm{eq}}\right),
\label{eq:RTA}
\end{equation}
where $F^{\alpha\beta}=\partial^\alpha A^\beta-\partial^\beta A^\alpha$.  We expand around a homogeneous equilibrium state with fixed temperature and fluid velocity,
\begin{equation}
f_0(\mathbf p)=\exp\!\left(\frac{\mu_0-p^0}{T_0}\right),\qquad
T=T_0,\qquad u^\mu=(1,\mathbf 0).
\label{eq:f0}
\end{equation}
The current is
\begin{equation}
J^\mu(x)=\int\frac{d^3p}{(2\pi)^3}\,v^\mu f(x,\mathbf p),\qquad v^\mu=(1,\mathbf v),
\label{eq:current}
\end{equation}
so that $n=J^0$.  Degeneracy factors can be restored by an overall multiplicative factor and will be suppressed.

We perturb the system with a weak gauge field $\delta A^\mu$ and allow the local chemical potential to respond, while keeping $T_0$ and $u^\mu$ fixed,
\begin{equation}
\mu(x)=\mu_0+\delta\mu(x).
\label{eq:thermo}
\end{equation}
Both the distribution and the chemical potential field are expanded in powers of the source,
\begin{align}
f&=f_0+\delta f^{(1)}+\delta f^{(2)}+\cdots,\\
\delta\mu&=\delta\mu^{(1)}+\delta\mu^{(2)}+\cdots,
\label{eq:expansion}
\end{align}
with $\delta f^{(n)}=O((\delta A)^n)$.  For a Boltzmann distribution,
\begin{equation}
f_{\mathrm{eq}}=f_0\exp\!\left(\frac{\delta\mu}{T_0}\right),
\label{eq:feq_expansion}
\end{equation}
so the second-order local-equilibrium contribution contains both $\delta\mu^{(2)}$ and $(\delta\mu^{(1)})^2/2T_0$.

The matching condition is chosen so that the RTA collision term does not violate charge conservation,
\begin{equation}
\int\frac{d^3p}{(2\pi)^3}\,[f-f_{\mathrm{eq}}]=0.
\label{eq:matching}
\end{equation}
Equivalently, the density computed from $f$ equals the density computed from the local-equilibrium distribution order by order.  This is the only conserved channel retained in the present setup; energy and momentum perturbations are frozen.  One may also directly implement this matching condition by constructing an  RTA kernel that explicitly recovers collision invariance \cite{Hu:2024tnn,Hu:2023elg,Lin:2025ehr}.

The fluctuation equation is
\begin{equation}
p^\mu\partial_\mu (f_0+\delta f)+\delta F^{i\beta}p_\beta\frac{\partial}{\partial p^i}(f_0+\delta f)=-\frac{p^0}{\tau_R}\left(\delta f-\delta f_{\mathrm{eq}}\right).
\label{eq:rta_fluctuation}
\end{equation}
We use
\begin{equation}
\widetilde Q(\omega,\mathbf k)=\int dt\,d^3\mathbf x\,e^{i(\omega t-\mathbf k\cdot\mathbf x)}Q(t,\mathbf x),\qquad
Q(t,\mathbf x)=\int\frac{d\omega\,d^3\mathbf k}{(2\pi)^4}e^{-i(\omega t-\mathbf k\cdot\mathbf x)}\widetilde Q(\omega,\mathbf k).
\label{eq:FT}
\end{equation}
At first order Eq.~\eqref{eq:rta_fluctuation} becomes
\begin{equation}
\left(\frac{1}{\tau_R}-i\omega+i\mathbf{k}\cdot\mathbf{v}\right)\delta f^{(1)}(\omega,\mathbf{k},\mathbf{p})=\frac{1}{\tau_R}\delta f^{(1)}_{\mathrm{eq}}(\omega,\mathbf{k},\mathbf{p})+\frac{f_0(\mathbf p)}{T_0}\mathbf v\cdot\mathbf E(\omega,\mathbf k),
\label{eq:rta_linear}
\end{equation}
with $\mathbf E$ the electric field generated by the perturbing potential.  Hence
\begin{equation}
\delta f^{(1)}(\omega,\mathbf{k},\mathbf{p}) = \frac{f_0}{T_0}\frac{\delta\mu^{(1)}(\omega,\mathbf{k}) + \tau_R\mathbf{v}\cdot\mathbf{E}}{1 + \tau_R(-i\omega + i\mathbf{k}\cdot\mathbf{v})}.
\label{eq:10}
\end{equation}
It is useful to introduce the streaming-relaxation operator
\begin{equation}
L\equiv \partial_t+\mathbf v\cdot\nabla+\frac{1}{\tau_R},\qquad
L^{-1}(\omega,\mathbf k)=\frac{1}{-i\omega+i\mathbf v\cdot\mathbf k+1/\tau_R}.
\label{eq:Lintro}
\end{equation}

\section{Linear response and the two-point function}
\label{sec:linear}

The matching condition gives
\begin{equation}
\delta n(\omega,\mathbf k)=\int\frac{d^3p}{(2\pi)^3}\delta f(\omega,\mathbf k,\mathbf p)
=\int\frac{d^3p}{(2\pi)^3}\delta f_{\mathrm{eq}}(\omega,\mathbf k,\mathbf p).
\label{deltan}
\end{equation}
At first order, $\delta n^{(1)}=\chi\,\delta\mu^{(1)}$ with
\begin{equation}
\chi\equiv\frac{\partial n_0}{\partial\mu_0}=\int\frac{dp}{2\pi^2T_0}\,p^2 f_0(p)=\frac{T_0^2}{\pi^2}e^{\mu_0/T_0}.
\label{chi}
\end{equation}
Substituting Eq.~\eqref{eq:10} into the matching condition yields the self-consistency equation
\begin{equation}
\delta n^{(1)}(\omega ,\mathbf{k}) = \chi \frac{\displaystyle\int\frac{d\Omega}{4\pi}\frac{\tau_R\mathbf{v}\cdot\mathbf{E}}{1 + \tau_R(-i\omega + i\mathbf{k}\cdot\mathbf{v})}}{\displaystyle 1 - \frac{1}{2ik\tau_R}\ln \left(\frac{\omega - k + \frac{i}{\tau_R}}{\omega + k + \frac{i}{\tau_R}}\right)},
\label{eq:13}
\end{equation}
where $k\equiv |\mathbf k|$.  Defining \footnote{Note that here we follow the convention of \cite{Romatschke:2015gic}, adopting a definition different from the traditional definition $G_R^{\mu,\nu}(K)\equiv\left.\frac{\delta J^\mu(K)}{\delta A_\nu(K)}\right|_{A=0}$. The same holds for Eq.\eqref{eq:threepoint_def}.}
\begin{equation}
G_R^{0,\nu}(K)\equiv\left.\frac{\delta n(K)}{\delta A^\nu(K)}\right|_{A=0},\qquad K=(\omega,\mathbf k),
\label{eq:linear_def}
\end{equation}
one obtains \cite{Romatschke:2015gic,Bajec:2024jez} 
\begin{align}
G_{R}^{0,0}(\omega ,\mathbf{k}) &= -\chi \frac{1 + (i\omega -\frac{1}{\tau_R})\frac{1}{2ik}\ln\left(\frac{\omega - k + \frac{i}{\tau_R}}{\omega + k + \frac{i}{\tau_R}}\right)}{1 - \frac{1}{2ik\tau_R}\ln\left(\frac{\omega - k + \frac{i}{\tau_R}}{\omega + k + \frac{i}{\tau_R}}\right)},
\label{eq:14a}\\
G_{R}^{0,i}(\omega ,\mathbf{k}) &= \frac{\omega\chi k^i}{k^2}\frac{1 + (i\omega -\frac{1}{\tau_R})\frac{1}{2ik}\ln\left(\frac{\omega - k + \frac{i}{\tau_R}}{\omega + k + \frac{i}{\tau_R}}\right)}{1 - \frac{1}{2ik\tau_R}\ln\left(\frac{\omega - k + \frac{i}{\tau_R}}{\omega + k + \frac{i}{\tau_R}}\right)}.
\label{eq:14b}
\end{align}
Other two-point correlators can also be calculated similarly. Charge conservation implies the Ward identity
\begin{equation}
-\omega G_{R}^{0,0}(\omega ,\mathbf{k}) = k_i G_{R}^{0,i}(\omega ,\mathbf{k}).
\label{eq:ward2}
\end{equation}
The logarithm has branch points at $\omega=\pm k-i/\tau_R$.  The same denominator also contains the hydrodynamic pole that, for $k\tau_R\ll1$, gives
\begin{equation}
\omega_{\mathrm{diff}}(k)=-iDk^2+O(k^4),\qquad D=\frac{\tau_R}{3}.
\label{eq:diffusion_pole}
\end{equation}
Discussions on two-point retarded correlation functions have been extensively carried out in \cite{Romatschke:2015gic, Kurkela:2017xis, Bajec:2025dqm, Hu:2024tnn,Lin:2025ehr,Zhao:2026mhx}. Hereafter, we focus on how to compute the three-point retarded correlation functions and discuss their physical properties.

\section{Nonlinear response and retarded three-point functions}
\label{sec:nonlinear}

The source expansion of the kinetic equation gives direct access to fully retarded higher-point response.  At order $n$, the only new unknown is $\delta f^{(n)}$; lower-order distributions act as known sources.  The $n$-th order kinetic equation is
\begin{equation}
p^\mu\partial_\mu \delta f^{(n)}+\delta F^{i\beta}p_\beta\frac{\partial}{\partial p^i}\delta f^{(n-1)}=-\frac{p^0}{\tau_R}\left(\delta f^{(n)}-\delta f_{\mathrm{eq}}^{(n)}\right).  \label{fn}
\end{equation}
Because $\delta f_{\mathrm{eq}}^{(n)}$ is fixed by matching and $\delta f^{(n-1)}$ is already known, Eq.~\eqref{fn} can be solved recursively as
\begin{align}
        \delta f^{(n)}(t,\mathbf{x},\mathbf{p})=\frac{1}{\partial_t+\mathbf{v}\cdot\nabla+\frac{1}{\tau_R}}\left(\frac{1}{\tau_R}\delta  f^{(n)}_{\mathrm{eq}}(t,\mathbf{x},\mathbf{p})-\delta F^{i\beta}v_\beta\frac{\partial}{\partial p^i}\delta f^{(n-1)}\right)
\label{eq:recursive}
\end{align}
formally.

We define the fully retarded three-point response by
\begin{equation}
G_R^{\mu,\nu\rho}(K_1,K_2)\equiv
\left.\frac{\delta^2 J^\mu(K_1+K_2)}{\delta A^\nu(K_1)\,\delta A^\rho(K_2)}\right|_{A=0}.
\label{eq:threepoint_def}
\end{equation}
The two source legs are therefore symmetric under $(\nu,K_1)\leftrightarrow(\rho,K_2)$.  In the usual Schwinger--Keldysh $r/a$ language this is a fully retarded $raa$-type correlator, up to convention-dependent factors.  We now specialize to $n=2$, the first nontrivial nonlinear response:
\begin{align}
        \delta f^{(2)}(t,\mathbf{x},\mathbf{p})=\frac{1}{\partial_t+\mathbf{v}\cdot\nabla+\frac{1}{\tau_R}}\Big(\frac{f_0}{2\tau_R}\frac{(\delta \mu^{(1)})^2}{ T_0^2}+\frac{f_0}{\tau_R}\frac{\delta \mu^{(2)}}{ T_0}-\delta F^{i\beta}v_\beta\frac{\partial}{\partial p^i}\big[\frac{1}{\partial_t+\mathbf{v}\cdot\nabla+\frac{1}{\tau_R}}\,(\frac{f_0\delta \mu^{(1)}}{\tau_R T_0}+\frac{f_0}{T_0}\mathbf{v}\cdot \mathbf{E}\,)\,\big] \Big). \label{f20}
\end{align}
Note that $L$ depends on momentum through $\mathbf v$, and its $p^i$ derivative is elementary.  Using $\partial_{p^i}L^{-1}=-L^{-1}(\partial_{p^i}L)L^{-1}$ gives
\begin{equation}
\frac{\partial L^{-1}}{\partial p^i}
= - \left[
\frac{(p^0)^2 \partial_i - p_i (\mathbf{p}\cdot\nabla)}{(p^0)^3}
\right] L^{-2}=- \frac{1}{p^0}\left(
\partial_i -v_i (\mathbf{v}\cdot\nabla)
\right) L^{-2}.
\end{equation}

Combining these identities with the first-order solution and the matching relation, Eq.~\eqref{f20} becomes
\begin{align}
        \delta f^{(2)}(t,\mathbf{x},\mathbf{p})
        &=\frac{f_0}{\tau_RT_0\chi}L^{-1}\delta n^{(2)}+\frac{f_0}{T_0\chi\tau_Rp^0}L^{-1}\Big(\,(\delta F^{i\beta}v_\beta)\nabla_iL^{-2} \delta n^{(1)}\,\Big)+\frac{f_0}{T_0p^0}L^{-1}\Big(\,(\delta F^{i\beta}v_\beta)\nabla_iL^{-2} \mathbf{v}\cdot \mathbf{E}\,\Big)\nn \\
        &-\frac{f_0}{T_0\chi\tau_Rp^0}L^{-1}\Big(\,(\delta F^{i\beta}v_\beta v_i)\mathbf{v}\cdot\nabla L^{-2} \delta n^{(1)}\,\Big)-\frac{f_0}{T_0p^0}L^{-1}\Big(\,(\delta F^{i\beta}v_\beta v_i)\mathbf{v}\cdot\nabla L^{-2} \mathbf{v}\cdot \mathbf{E}\,\Big)\nn \\
        &-\frac{f_0}{T_0p^0}L^{-1}\big(\,(\delta F^{i\beta}v_\beta)L^{-1}E_i\big)+\frac{f_0}{T_0p^0}L^{-1}\big(\,(\delta F^{i\beta}v_\beta v_i)L^{-1}\mathbf{v}\cdot \mathbf{E}\big)\nn\\
        &+\frac{f_0}{\chi T^2_0\tau_R}L^{-1}\big(\delta F^{i\beta}v_\beta v_iL^{-1} \delta n^{(1)}\big)+\frac{f_0}{T^2_0}L^{-1}\big(\delta F^{i\beta}v_\beta v_iL^{-1}\mathbf{v}\cdot \mathbf{E}\big),
        \label{f21}
\end{align}
where 
\begin{align}
 \delta \mu^{(2)}&=\frac{1}{\chi} \delta n^{(2)}-\frac{1}{2T_0\chi^2} (\delta n^{(1)})^2
\label{eq:mu2}
\end{align}
is utilized.

Integrating Eq.~\eqref{f21} over momentum and imposing the matching condition produces a closed equation for $\delta n^{(2)}$:
\begin{align}
        \delta n^{(2)}(t,\mathbf{x})&=\int \frac{d^3p}{(2\pi)^3} \Big(\frac{f_0}{\tau_RT_0\chi}L^{-1}\delta n^{(2)}+\frac{f_0}{T_0\chi\tau_Rp^0}L^{-1}\big(\,(\delta F^{i\beta}v_\beta)\nabla_iL^{-2} \delta n^{(1)}\,\big)+\frac{f_0}{T_0p^0}L^{-1}\big(\,(\delta F^{i\beta}v_\beta)\nabla_iL^{-2} \mathbf{v}\cdot \mathbf{E}\,\big)\nn \\
        &-\frac{f_0}{T_0\chi\tau_Rp^0}L^{-1}\big(\,(\delta F^{i\beta}v_\beta v_i)\mathbf{v}\cdot\nabla L^{-2} \delta n^{(1)}\,\big)-\frac{f_0}{T_0p^0}L^{-1}\big(\,(\delta F^{i\beta}v_\beta v_i)\mathbf{v}\cdot\nabla L^{-2} \mathbf{v}\cdot \mathbf{E}\,\big)\nn \\
        &-\frac{f_0}{T_0p^0}L^{-1}\big(\,(\delta F^{i\beta}v_\beta)L^{-1}E_i\big)+\frac{f_0}{T_0p^0}L^{-1}\big(\,(\delta F^{i\beta}v_\beta v_i)L^{-1}\mathbf{v}\cdot \mathbf{E}\big)\nn\\
        &+\frac{f_0}{\chi T^2_0\tau_R}L^{-1}\big(\delta F^{i\beta}v_\beta v_iL^{-1} \delta n^{(1)}\big)+\frac{f_0}{T^2_0}L^{-1}\big(\delta F^{i\beta}v_\beta v_iL^{-1}\mathbf{v}\cdot \mathbf{E}\big)\,\Big).
        \label{deltan2}
\end{align}
After Fourier transforming and taking two functional derivatives with respect to the scalar source, we obtain
\begin{align}
    G_R^{0,00}&(\omega-\omega^{\prime},\mathbf{k}-\mathbf{k}^{\prime};\omega^{\prime},\mathbf{k}^{\prime})=\frac{1}{1 - \frac{1}{2i|\mathbf{k}|\tau_R}\ln \left(\frac{\omega - |\mathbf{k}| + \frac{i}{\tau_R}}{\omega + |\mathbf{k}| + \frac{i}{\tau_R}}\right)}\int \frac{d^3p}{(2\pi)^3}\frac{f_0}{T_0p^0}\nn\\
        &\times  \Big(\frac{1}{\chi\tau_R}L^{-1}(\omega,\mathbf{k})\big(\,(\mathbf{k}-\mathbf{k}^{\prime})\cdot \mathbf{k}^{\prime}L^{-2}(\omega^{\prime},\mathbf{k}^{\prime}) G_R^{0,0}(\omega^{\prime},\mathbf{k}^{\prime})\,\big)\nn\\
        &+L^{-1}(\omega,\mathbf{k})\big(\,(\mathbf{k}-\mathbf{k}^{\prime})\cdot \mathbf{k}^{\prime} (-i\mathbf{v}\cdot\mathbf{k}^{\prime})L^{-2}(\omega^{\prime},\mathbf{k}^{\prime})\,\big)\nn \\
        &-\frac{1}{\chi\tau_R}L^{-1}(\omega,\mathbf{k})\big(\,\mathbf{v}\cdot(\mathbf{k}-\mathbf{k}^{\prime})(\mathbf{v}\cdot\mathbf{k}^{\prime})L^{-2}(\omega^{\prime},\mathbf{k}^{\prime}) G^{0,0}_R (\omega^{\prime},\mathbf{k}^{\prime})\,\big)\nn\\
        &-L^{-1}(\omega,\mathbf{k})\big(\,-i\mathbf{v}\cdot(\mathbf{k}-\mathbf{k}^{\prime})(\mathbf{v}\cdot\mathbf{k}^{\prime})^2L^{-2}(\omega^{\prime},\mathbf{k}^{\prime}) \,\big)\nn\\
        &-L^{-1}(\omega,\mathbf{k})\big(\,-(\mathbf{k}-\mathbf{k}^{\prime})\cdot \mathbf{k}^{\prime} L^{-1}(\omega^{\prime},\mathbf{k}^{\prime})\big)\,\nn\\
        &+L^{-1}(\omega,\mathbf{k})\big(\,-\mathbf{v}\cdot(\mathbf{k}-\mathbf{k}^{\prime})(\mathbf{v}\cdot\mathbf{k}^{\prime})L^{-1}(\omega^{\prime},\mathbf{k}^{\prime}) \,\big)\nn\\
        &+\frac{p^0}{\chi T_0\tau_R}L^{-1}(\omega,\mathbf{k})\big(\,-i\mathbf{v}\cdot(\mathbf{k}-\mathbf{k}^{\prime})L^{-1}(\omega^{\prime},\mathbf{k}^{\prime})G_R^{0,0}(\omega^{\prime},\mathbf{k}^{\prime}) \,\big)\nn\\
        &+\frac{p^0}{T_0}L^{-1}(\omega,\mathbf{k})\big(\,-\mathbf{v}\cdot(\mathbf{k}-\mathbf{k}^{\prime})(\mathbf{v}\cdot\mathbf{k}^{\prime})L^{-1}(\omega^{\prime},\mathbf{k}^{\prime}) \,\big)\nn\\
        &+[\omega-\omega^{\prime}\leftrightarrow \omega^{\prime}, \mathbf{k}-\mathbf{k}^{\prime}\leftrightarrow\mathbf{k}^{\prime}] \Big)\label{GJ000}.
\end{align}

The mixed temporal--spatial components $G_R^{0,0i}$ and $G_R^{0,ij}$ are collected in Appendix~\ref{correlator}.  Together they satisfy the three-point Ward identities
\begin{align}
    &\omega^\prime    G_R^{0,00}(\omega-\omega^{\prime},\mathbf{k}-\mathbf{k}^{\prime};\omega^{\prime},\mathbf{k}^{\prime}) + k_i^\prime G_R^{0,0i}(\omega-\omega^{\prime},\mathbf{k}-\mathbf{k}^{\prime};\omega^{\prime},\mathbf{k}^{\prime})=0, \\
    &(\omega-\omega^\prime )    G_R^{0,0j}(\omega-\omega^{\prime},\mathbf{k}-\mathbf{k}^{\prime};\omega^{\prime},\mathbf{k}^{\prime}) + (k_i-k_i^\prime )G_R^{0,ij}(\omega-\omega^{\prime},\mathbf{k}-\mathbf{k}^{\prime};\omega^{\prime},\mathbf{k}^{\prime})=0,
\label{eq:ward3}
\end{align}
which provide several useful checks on the lengthy three-point expression up to now.

For the remainder of the main text we focus on the density response $G_R^{0,00}$.  The radial momentum integrals factorize from the angular dependence.  Define
\begin{align}
    F(\omega,\omega^{\prime},\mathbf{k},\mathbf{k}^\prime,m,n,l,h)\equiv  \int \frac{d^3p}{(2\pi)^3} \frac{f_0}{(p^0)^h}(\mathbf{v}\cdot\mathbf{k})^m(\mathbf{v}\cdot\mathbf{k}^\prime)^nL^{-1}(\omega,\mathbf{k})L^{-l}(\omega^{\prime},\mathbf{k}^{\prime}),
\label{eq:F}
\end{align}
then
\begin{align}
 &F(\omega,\omega^{\prime},\mathbf{k},\mathbf{k}^\prime,m,n,l,1)=\frac{T_0^2 e^{\mu_0/T_0}}{4 \pi ^2} I_{mnl}(\omega,\omega^{\prime},\mathbf{k},\mathbf{k}^\prime), \\
&F(\omega,\omega^{\prime},\mathbf{k},\mathbf{k}^\prime,m,n,l,0)=\frac{T_0^3 e^{\mu_0/T_0}}{2\pi ^2} I_{mnl}(\omega,\omega^{\prime},\mathbf{k},\mathbf{k}^\prime),
\end{align}
with the general angular integral defined as
\[
 I_{mnl}\equiv \int \frac{d\Omega}{2\pi}\frac{(\mathbf{v}\cdot\mathbf{k})^m(\mathbf{v}\cdot\mathbf{k}^\prime)^n}{-i\omega+i\mathbf{v}\cdot\mathbf{k}+\frac{1}{\tau_R}}\frac{1}{(-i\omega^\prime+i\mathbf{v}\cdot\mathbf{k}^\prime+\frac{1}{\tau_R})^l}.
\]
 The required angular integrals are evaluated and given in the appendices.  In this notation the density three-point function is
\begin{align}
   & G_R^{0,00}(\omega-\omega^{\prime},\mathbf{k}-\mathbf{k}^{\prime};\omega^{\prime},\mathbf{k}^{\prime})=\frac{1}{1 - \frac{1}{2i|\mathbf{k}|\tau_R}\ln \left(\frac{\omega - |\mathbf{k}| + \frac{i}{\tau_R}}{\omega + |\mathbf{k}| + \frac{i}{\tau_R}}\right)}\nn\\
        &\times  \Big(\frac{1}{T_0\chi\tau_R}\,(\mathbf{k}-\mathbf{k}^{\prime})\cdot \mathbf{k}^{\prime}F(\omega,\omega^{\prime},\mathbf{k},\mathbf{k}^\prime,0,0,2,1)  G_R^{0,0}(\omega^{\prime},\mathbf{k}^{\prime})\,\nn\\
        &+\frac{1}{T_0}\,(\mathbf{k}-\mathbf{k}^{\prime})\cdot \mathbf{k}^{\prime}  (-i)F(\omega,\omega^{\prime},\mathbf{k},\mathbf{k}^\prime,0,1,2,1)\nn \\
        &-\frac{1}{T_0\chi\tau_R}\big(\,F(\omega,\omega^{\prime},\mathbf{k},\mathbf{k}^\prime,1,1,2,1)-F(\omega,\omega^{\prime},\mathbf{k},\mathbf{k}^\prime,0,2,2,1) \,\big)G^{0,0}_R (\omega^{\prime},\mathbf{k}^{\prime})\nn\\
        &-\frac{1}{T_0}(-i)\big( F(\omega,\omega^{\prime},\mathbf{k},\mathbf{k}^\prime,1,2,2,1)-F(\omega,\omega^{\prime},\mathbf{k},\mathbf{k}^\prime,0,3,2,1)\big) \nn\\
        &-\frac{1}{T_0}\big(-(\mathbf{k}-\mathbf{k}^{\prime})\cdot \mathbf{k}^{\prime} \big)F(\omega,\omega^{\prime},\mathbf{k},\mathbf{k}^\prime,0,0,1,1)\,\nn\\
        &+\frac{1}{T_0}(-1)\big(F(\omega,\omega^{\prime},\mathbf{k},\mathbf{k}^\prime,1,1,1,1)-F(\omega,\omega^{\prime},\mathbf{k},\mathbf{k}^\prime,0,2,1,1)\big)\nn\\
        &+\frac{1}{\chi T_0^2\tau_R}(-i)\big(\,F(\omega,\omega^{\prime},\mathbf{k},\mathbf{k}^\prime,1,0,1,0)-F(\omega,\omega^{\prime},\mathbf{k},\mathbf{k}^\prime,0,1,1,0) \,\big)G_R^{0,0}(\omega^{\prime},\mathbf{k}^{\prime})\nn\\
        &+\frac{1}{T_0^2}(-1)\big(\,F(\omega,\omega^{\prime},\mathbf{k},\mathbf{k}^\prime,1,1,1,0)-F(\omega,\omega^{\prime},\mathbf{k},\mathbf{k}^\prime,0,2,1,0)\,\big)\nn\\
        &+[\omega-\omega^{\prime}\leftrightarrow \omega^{\prime}, \mathbf{k}-\mathbf{k}^{\prime}\leftrightarrow\mathbf{k}^{\prime}] \Big),
        \label{GJ0001}
\end{align}
where all the $I_{nml}$ appearing in the above expressions are computed and given in the Appendix; they are all composed of elementary functions. The final term explicitly enforces the exchange symmetry of the two source legs.  This symmetry is a direct consequence of the functional-derivative definition in Eq.~\eqref{eq:threepoint_def}.

\section{Analytic structure of the three-point response}
\label{sec:analytic}

Momentum conservation assigns the three external legs of $G_R^{0,00}$ the four-momenta
\begin{equation}
K_1=(\omega-\omega',\mathbf k-\mathbf k'),\qquad
K_2=(\omega',\mathbf k'),\qquad
K_3=(-\omega,-\mathbf k),\qquad
\qquad K_1+K_2+K_3=0.
\label{eq:legs}
\end{equation}
The analytic structure can be separated into structures inherited from individual legs and structures generated by products of kinetic propagators.

\subsection{External-leg singularities}

The overall  denominator in Eq.~\eqref{GJ0001} of the form 
\begin{align}
    \frac{1}{1 - \frac{1}{2i|\mathbf{k}|\tau_R}\ln \left(\frac{\omega - |\mathbf{k}| + \frac{i}{\tau_R}}{\omega + |\mathbf{k}| + \frac{i}{\tau_R}}\right)}
\end{align}
is the same function that appears in the two-point response.  It has logarithmic branch points at
\begin{equation}
\omega=\pm |\mathbf k|-\frac{i}{\tau_R}.
\label{leg1}
\end{equation}
The two-point functions nested inside the three-point expression produce the corresponding structures on the two source legs,
\begin{align}
\omega'&=\pm |\mathbf k'|-\frac{i}{\tau_R},
\label{leg2}\\
\omega-\omega'&=\pm |\mathbf k-\mathbf k'|-\frac{i}{\tau_R}.
\label{leg3}
\end{align}
Thus the nonlinear response necessarily remembers the same analytical structure that governs linear response, now replicated on all three external legs.

\subsection{Mixed kinetic singularities}

The genuinely new kinematics enters through angular integrals containing more than one streaming denominator.  The simplest example is
\begin{equation}
I_{001}=\int\frac{d\Omega}{2\pi}\frac{1}{(a+i\mathbf v\cdot\mathbf k)(b+i\mathbf v\cdot\mathbf k')},
\qquad a=\frac1{\tau_R}-i\omega,\quad b=\frac1{\tau_R}-i\omega'.
\label{eq:I001main}
\end{equation}
$I_{002}$ is related to it through
derivative, so here we focus on the analysis of $I_{001}$.
As shown in Appendix~\ref{app:I001}, Feynman parametrization reduces this integral to a quadratic denominator $D(t)=D_2t^2+D_1t+D_0$.  Its closed form contains
\begin{equation}
\sqrt{4D_0D_2-D_1^2},
\end{equation}
so the candidate branch points of square-root type could  occur when
\begin{equation}
D_1^2-4D_0D_2=0.
\label{eq:mixed_discriminant}
\end{equation}
Equivalently,
\begin{equation}
D_1^2-4D_0D_2=-4\left[a^2k'^2-2ab\,\mathbf k\!\cdot\!\mathbf k'+b^2k^2+k^2k'^2-(\mathbf k\!\cdot\!\mathbf k')^2\right]=0.
\label{eq:mixed_explicit}
\end{equation}
This condition mixes the frequencies and momenta of different legs and therefore cannot be assigned to a single external propagator.  In kinetic terms it is a Landau-type threshold: two streaming denominators can become resonant for the same particle direction. Once this condition is satisfied, the angular integral develops a pinch singularity, which in turn generates a branch point in the three-point function. Here we write down the equations that the branch points must satisfy, but we do not solve them explicitly
because of the complexity of the resulting expressions. 

In addition, there are also  logarithmic branch points decided by $D_0=0$ and $D_0+D_1+D_2=0$ because of $\arctan$ function, which are found to coincide with  the ordinary single-leg branch points in Eqs.~\eqref{leg2} and \eqref{leg1}, respectively. We further note that the singularity structures of the remaining $I_{nml}$ agree with Eqs.~\eqref{leg1}, \eqref{leg2}, and \eqref{leg3}, except that $M_{102}$ (see Eq.\eqref{M102}) yields poles described by \eqref{leg2} and \eqref{leg3}, i.e., the branch point and the pole collide.


\section{Consistency limits and physical interpretation}
\label{sec:limits}

  It is manifest that $G_R^{0,00}(\omega-\omega^{\prime},\mathbf{k}-\mathbf{k}^{\prime};\omega^{\prime},\mathbf{k}^{\prime})$ vanishes in the limits $\mathbf{k}^{\prime}\to 0$ and $\mathbf{k}-\mathbf{k}^{\prime}\to 0$, as required by charge conservation. The limit $\mathbf{k}\to 0$, however, is less straightforward to verify, because the expression for $G_R^{0,00}$ is complicated and contains many terms. Nevertheless, one can still prove analytically that
  \begin{align}
    \lim_{\mathbf{k}\to 0}  G_R^{0,00}(\omega-\omega^{\prime},\mathbf{k}-\mathbf{k}^{\prime};\omega^{\prime},\mathbf{k}^{\prime})\to 0,
  \end{align}
   which serves as a highly nontrivial consistency check.
  
  On the other hand, a spatially homogeneous time-dependent scalar potential is pure gauge and cannot change the density.  The strictly static limit is subtler because a constant $A^0$ shifts the thermodynamic chemical potential; consequently the limit $\omega'\to0$  followed by $\mathbf k'\to0$ is not interchangeable with a generic dynamical zero-momentum limit followed by $\omega'\to0$. Taking the source leg static first and then homogeneous gives
\begin{equation}
\lim_{\mathbf{k}^{\prime}\to 0}\lim_{\omega^{\prime}\to 0}
G_R^{0,00}(\omega-\omega^{\prime},\mathbf{k}-\mathbf{k}^{\prime};\omega^{\prime},\mathbf{k}^{\prime})
=-\frac{1}{T_0}G_R^{0,0}(\omega,\mathbf{k})
=-\frac{\partial}{\partial \mu_0}G_R^{0,0}(\omega,\mathbf{k}).
\label{eq:static_limit}
\end{equation}
The last equality follows from the Boltzmann-gas dependence $\chi\propto e^{\mu_0/T_0}$.  Equation~\eqref{eq:static_limit} is the expected thermodynamic insertion relation: a zero-frequency, zero-momentum density insertion measures the dependence of the lower-point response on the background chemical potential.  It therefore supplies a nontrivial consistency check on the nonlinear calculation.

A second useful limit is $\tau_R\to\infty$. In this algebraic limit we find an enlarged permutation
property among the external legs. To be specific, we examine whether the three-point function is symmetric under the exchange of the arguments associated with its 2,3 or 1,3 external legs. The exchange symmetry between legs 1 and 2 is manifest. To test the other exchanges, we consider the difference
\begin{align}
 G_R^{0,00}(\omega-\omega^{\prime},\mathbf{k}-\mathbf{k}^{\prime};\omega^{\prime},\mathbf{k}^{\prime})- G_R^{0,00}(-\omega,-\mathbf{k};\omega^{\prime},\mathbf{k}^{\prime}),
\end{align}
where the third leg is implicit and fixed by energy-momentum conservation. The second term corresponds to exchanging legs 1 and 3, which is equivalent to the replacement $(\omega,\mathbf{k})\rightarrow (\omega^\prime-\omega,\mathbf{k}^\prime-\mathbf{k})$. In general, this difference does not vanish. However, in the collisionless limit $\tau_R\to\infty$, it vanishes identically away from the resonance points (the singularities of the correlation functions):
\begin{align}
 \lim_{\tau_R\to \infty}\big( G_R^{0,00}(\omega-\omega^{\prime},\mathbf{k}-\mathbf{k}^{\prime};\omega^{\prime},\mathbf{k}^{\prime})- G_R^{0,00}(-\omega,-\mathbf{k};\omega^{\prime},\mathbf{k}^{\prime})\big) \to 0, \quad \text{away from the resonance region}.
\end{align}
At the singularities, setting $1/\tau_R$ exactly to zero is not legitimate, and the order in which this limit is taken matters. To avoid the divergence, $1/\tau_R$ must be given an infinitesimal positive real part, $1/\tau_R\to 0^{+}$. In the field-theoretic prescription, $0^{+}$ selects the causal retarded response, which differs from genuinely collisionless dynamics.  
Away from the resonance region, $1/\tau_R$ can be set exactly to zero, recovering the collisionless Vlasov equation. In this limit, the difference vanishes \cite{Delacretaz:2022ocm}. We now discuss the physical meaning of this result from the perspective of correlation functions. In the $r$-$a$ basis of non-equilibrium field theory, the three-point function obtained here is of the $G^{raa}$ type\footnote{Below, we omit the $0$ component signature when no confusion arises.}, with one $r$-type leg and two $a$-type legs. It arises from the second variation of an $r$-type physical field with respect to the $a$-type sources. Therefore, the exchange symmetry should be restricted to the frequencies and wavevectors of the two $a$-type legs, as is explicit in Eq.~\eqref{GJ0001}. In the collisionless limit, there is no dissipation and hence no arrow of time, so the notion of causal response loses its meaning. Consequently, the distinction between $r$ and $a$ fields is no longer sharp; indeed, the $r$-$a$ basis was originally introduced to describe dissipative phenomena in non-equilibrium field theory, such as the Schwinger-Keldysh formalism, and becomes unnecessary in this limit. Thus, the frequency and wavevector carried by the $r$-type leg  exhibit an enlarged exchange symmetry with those of the $a$-type legs, similar to the collisionless case \cite{Delacretaz:2022ocm}. 



\section{Relation to Schwinger--Keldysh effective theory}
\label{sec:comparison}

In Ref.~\cite{Abbasi:2024pwz}, the authors construct a Schwinger-Keldysh effective field theory (EFT) for density diffusion to all orders in derivatives. Starting from the Boltzmann equation in the relaxation time approximation and employing a suitable derivative counting scheme, they obtain a resummation of the diffusion equation. The linear diffusion equation obtained from kinetic theory is then used as the input for the free part of the effective action, which is constructed through the Martin-Siggia-Rose (MSR) formalism \cite{Martin:1973zz}, while the fluctuation part is fixed by the dynamical KMS symmetry \cite{Crossley:2015evo,Glorioso:2017fpd}. Promoting the Wilsonian coefficients to functions of the dynamical field and expanding them, they also construct the interacting part of the effective action. Based on this effective action, they compute the symmetrized three-point correlation function $G^{rrr}$ for the charge density.

Below we compare our work with Ref.~\cite{Abbasi:2024pwz}, focusing on their connections and differences. First, both works share the same basic setup, i.e., the Boltzmann equation in the relaxation time approximation. However, our approach to computing the three-point function is more straightforward and relies on fewer approximations and assumptions than the EFT construction of Ref.~\cite{Abbasi:2024pwz}. Clearly, our present calculation can be directly and efficiently adapted to the study of momentum transport, or of combined charge and momentum transport, whereas this is not the case for \cite{Abbasi:2024pwz}.  On the other hand, because the Boltzmann equation does not incorporate the degrees of freedom needed to describe fluctuations i.e., it does not couple to an $a$-type source, the three-point functions we can compute are necessarily of the retarded type, $G^{raa}$.  By contrast, the framework of Ref.~\cite{Abbasi:2024pwz} is free from this limitation and can, in principle, compute various types of three-point functions, although only $G^{rrr}$ was explicitly computed there. Since the KMS condition for three-point functions does not provide a direct one-to-one map between $G^{raa}$ and $G^{rrr}$ \cite{Wang:1998wg}, a direct comparison of the three-point functions obtained in the two works is not readily available  at this stage. As mentioned above, however, the framework of Ref.~\cite{Abbasi:2024pwz} can be used to compute $G^{raa}$ directly, in which case the results presented here would serve as a useful benchmark.


On the other hand, once the three-point function is known, a natural question arises: can it be used as an interaction vertex to compute corrections to the two-point function, thereby yielding fluctuation corrections to transport coefficients or the relaxation time ? In Ref.~\cite{Abbasi:2024pwz}, although the corrected two-point function was not fully worked out, the one-loop correction to the analytic structure of the charge-density response function was computed. We may ask whether a similar calculation can be carried out using only the results obtained here. The answer is negative. At this stage, we have the two-point retarded correlation function; $G^{rr}$ follows from the fluctuation-dissipation theorem
\begin{equation}
G^{rr}(\omega,\mathbf{k})=-\frac{2T_0}{\omega}\,\mathrm{Im}\,G^{ra}(\omega,\mathbf{k})
\label{eq:FDT}
\end{equation}
in the classical limit, while $G^{ar}=(G^{ra})^*$. Given these two-point correlators and $G^{raa}$, any one-loop correction to the two-point function must involve an internal line of $G^{aa}$, which vanishes identically. In other words, with only retarded-type vertices available, fluctuation corrections to the two-point function cannot be computed. This is physically reasonable, as it reflects the fact that the Boltzmann equation does not capture fluctuations.

This distinction also limits what can be inferred from the deterministic Boltzmann equation alone.  Response vertices do not specify the independent stochastic vertices or higher noise cumulants of the effective theory.  Consequently, fluctuation-loop corrections to the two-point function cannot be closed using only the response functions derived here.  Such a calculation requires either a Boltzmann--Langevin description with specified noise statistics or a Schwinger--Keldysh effective action satisfying the appropriate KMS constraints.  Conversely, the explicit $G_R^{0,00}$ obtained here provides a microscopic benchmark that a kinetic Schwinger--Keldysh theory, such as the one developed in \cite{Abbasi:2024pwz}, should reproduce in its fully retarded sector.

\section{Summary and outlook}
\label{sec:summary}

In this paper, we have studied nonlinear response and three-point retarded correlation functions of charge transport in the relaxation time approximation of kinetic theory. Starting from the relativistic Boltzmann equation in the presence of an external $U(1)$ gauge field, we derived the $n$-th order response of the single-particle distribution function and analytically computed the three-point retarded correlation functions for the charge density  at generic frequencies and momenta. We analyzed the analytical structure of the three-point function and identified two classes of singularities: those inherited from the two-point function, and new structures arising from the mixing of different external legs. We verified that the correlation functions satisfy the Ward identities associated with charge conservation and discussed their behavior in various limits. We further compared our results with those obtained from the Schwinger-Keldysh effective field theory approach and discussed the limitations of the kinetic-theory framework.

Several extensions are natural.  The first is to construct the complete Schwinger--Keldysh kinetic theory that reproduces the retarded vertices derived here while simultaneously fixing the noise sector through KMS symmetry.  This would permit a direct comparison of $G^{raa}$ and $G^{rrr}$ and make fluctuation-loop corrections calculable.  A second direction is to replace the constant relaxation time by momentum-dependent or more realistic collision operators, for which the non-hydrodynamic spectrum may become gapless and the mixed singularities should change qualitatively.  Third, the same response expansion can be generalized to finite mass, multiple conserved charges \cite{Hu:2022vph}, background fields, and even with spin degrees of freedom \cite{Hu:2022lpi,Hu:2022xjn,Hu:2022mvl}, where anisotropy and additional slow modes enrich the analytic structure of three point  correlators.  


\vspace{1cm}
\noindent\textbf{Acknowledgments.} This work was supported by the National Natural Science Foundation of China under Grant No.~12505149. DeepSeek's DeepSeek-V4-Flash was helpful in editing and proofreading  the manuscript. 

\appendix

\section{The explicit form of \texorpdfstring{$G^{0,0i}_R$ and $G^{0,ij}_R$}{the remaining three-point components}}
\label{correlator}

For completeness, this appendix gives the remaining tensor components needed to verify the nonlinear Ward identities.  Taking the $A^0A^i$ derivatives of Eq.~\eqref{deltan2} gives
\begin{align}
    G_R^{0,0i}&(\omega-\omega^{\prime},\mathbf{k}-\mathbf{k}^{\prime};\omega^{\prime},\mathbf{k}^{\prime})=\frac{1}{1 - \frac{1}{2i|\mathbf{k}|\tau_R}\ln \left(\frac{\omega - |\mathbf{k}| + \frac{i}{\tau_R}}{\omega + |\mathbf{k}| + \frac{i}{\tau_R}}\right)}\int \frac{d^3p}{(2\pi)^3}\frac{f_0}{T_0p^0}\nn\\
        &\times  \Big(\frac{1}{\chi\tau_R}L^{-1}(\omega,\mathbf{k})\big(\,(\mathbf{k}-\mathbf{k}^{\prime})\cdot \mathbf{k}^{\prime}L^{-2}(\omega^{\prime},\mathbf{k}^{\prime}) G_R^{0,i}(\omega^{\prime},\mathbf{k}^{\prime})\nn\\
        &-\omega^{\prime} (k_i-k_i^{\prime})L^{-2}(\omega-\omega^{\prime},\mathbf{k}-\mathbf{k}^{\prime}) G_R^{0,0}(\omega-\omega^{\prime},\mathbf{k}-\mathbf{k}^{\prime})\,\big)\nn\\
        &+L^{-1}(\omega,\mathbf{k})\big(\,(\mathbf{k}-\mathbf{k}^{\prime})\cdot \mathbf{k}^{\prime}L^{-2}(\omega^{\prime},\mathbf{k}^{\prime}) (iv_i\omega^{\prime})+\omega^{\prime} (k_i-k_i^{\prime})L^{-2}(\omega-\omega^{\prime},\mathbf{k}-\mathbf{k}^{\prime}) (i\mathbf{v}\cdot(\mathbf{k}-\mathbf{k}^{\prime})\,)\,\big)\nn \\
        &-\frac{1}{\chi\tau_R}L^{-1}(\omega,\mathbf{k})\big(\,\mathbf{v}\cdot(\mathbf{k}-\mathbf{k}^{\prime})\mathbf{v}\cdot\mathbf{k}^{\prime}L^{-2}(\omega^{\prime},\mathbf{k}^{\prime}) G^{0,i}_R (\omega^{\prime},\mathbf{k}^{\prime})\nn\\
        &\,-v_i\omega^{\prime}\mathbf{v}\cdot(\mathbf{k}-\mathbf{k}^{\prime})L^{-2}(\omega-\omega^{\prime},\mathbf{k}-\mathbf{k}^{\prime}) G^{0,0}_R (\omega-\omega^{\prime},\mathbf{k}-\mathbf{k}^{\prime})\,\big)\nn\\
        &-L^{-1}(\omega,\mathbf{k})\big(v_i\omega^\prime\mathbf{v}\cdot(\mathbf{k}-\mathbf{k}^{\prime})(i\mathbf{v}\cdot\mathbf{k}^{\prime})L^{-2}(\omega^{\prime},\mathbf{k}^{\prime})\nn\\
        &+iv_i\omega^\prime\mathbf{v}\cdot(\mathbf{k}-\mathbf{k}^{\prime})\mathbf{v}\cdot(\mathbf{k}-\mathbf{k}^{\prime})L^{-2}(\omega-\omega^{\prime},\mathbf{k}-\mathbf{k}^{\prime})\big)\nn\\
        &-L^{-1}(\omega,\mathbf{k})\big(\,\omega^\prime(k_i-k_i^\prime)L^{-1}(\omega^{\prime},\mathbf{k}^{\prime})+\omega^\prime(k_i-k_i^\prime)L^{-1}(\omega-\omega^{\prime},\mathbf{k}-\mathbf{k}^{\prime}) \big)\nn\\
        &-L^{-1}(\omega,\mathbf{k})\big(\,v_i(\mathbf{k}-\mathbf{k}^{\prime})\cdot \mathbf{k}^{\prime}L^{-1}(\omega^{\prime},\mathbf{k}^{\prime})-(k_i-k_i^{\prime})\mathbf{v}\cdot\mathbf{k}^{\prime} L^{-1}(\omega-\omega^{\prime},\mathbf{k}-\mathbf{k}^{\prime}) \,\big)\nn \\
        &+L^{-1}(\omega,\mathbf{k})\big(v_i\omega^\prime\mathbf{v}\cdot(\mathbf{k}-\mathbf{k}^{\prime})L^{-1}(\omega^{\prime},\mathbf{k}^{\prime})+v_i\omega^\prime\mathbf{v}\cdot(\mathbf{k}-\mathbf{k}^{\prime})L^{-1}(\omega-\omega^{\prime},\mathbf{k}-\mathbf{k}^{\prime})\big) \nn\\
        &+\frac{p^0}{\chi T_0\tau_R}L^{-1}(\omega,\mathbf{k})\big(-i\mathbf{v}\cdot(\mathbf{k}-\mathbf{k}^{\prime})L^{-1}(\omega^{\prime},\mathbf{k}^{\prime})G_R^{0,i}(\omega^{\prime},\mathbf{k}^{\prime})\nn\\
        &+iv_i\omega^\prime L^{-1}(\omega-\omega^{\prime},\mathbf{k}-\mathbf{k}^{\prime})G^{0,0}_R (\omega-\omega^{\prime},\mathbf{k}-\mathbf{k}^{\prime})\,\big) \nn\\
        &+\frac{p^0}{T_0}L^{-1}(\omega,\mathbf{k})\big(v_i\omega^\prime\mathbf{v}\cdot(\mathbf{k}-\mathbf{k}^{\prime})L^{-1}(\omega^{\prime},\mathbf{k}^{\prime})+v_i\omega^\prime\mathbf{v}\cdot(\mathbf{k}-\mathbf{k}^{\prime})L^{-1}(\omega-\omega^{\prime},\mathbf{k}-\mathbf{k}^{\prime})\big)
        \,\Big).
        \label{GJ003}
\end{align}

The component with two spatial source indices is obtained analogously:
\begin{align}
    G_R^{0,ij}&(\omega-\omega^{\prime},\mathbf{k}-\mathbf{k}^{\prime};\omega^{\prime},\mathbf{k}^{\prime})=\frac{1}{1 - \frac{1}{2i|\mathbf{k}|\tau_R}\ln \left(\frac{\omega - |\mathbf{k}| + \frac{i}{\tau_R}}{\omega + |\mathbf{k}| + \frac{i}{\tau_R}}\right)}\int \frac{d^3p}{(2\pi)^3}\frac{f_0}{T_0p^0}\nn\\
        &\times  \Big(\frac{1}{\chi\tau_R}L^{-1}(\omega,\mathbf{k})\big(\,-(\omega-\omega^{\prime}) k_i^{\prime}L^{-2}(\omega^{\prime},\mathbf{k}^{\prime}) G_R^{0,j}(\omega^{\prime},\mathbf{k}^{\prime})\,\big)\nn\\
        &+L^{-1}(\omega,\mathbf{k})\big(\,-i\omega^{\prime}(\omega-\omega^{\prime}) k_i^{\prime}v_jL^{-2}(\omega^{\prime},\mathbf{k}^{\prime}) \,\big)\nn \\
        &-\frac{1}{\chi\tau_R}L^{-1}(\omega,\mathbf{k})\big(-(\omega-\omega^{\prime})v_i(\mathbf{v}\cdot\mathbf{k}^{\prime})L^{-2}(\omega^{\prime},\mathbf{k}^{\prime}) G^{0,j}_R (\omega^{\prime},\mathbf{k}^{\prime})\,\big)\nn\\
        &-L^{-1}(\omega,\mathbf{k})\big(\,-iv_iv_j\omega^{\prime}(\omega-\omega^{\prime})(\mathbf{v}\cdot\mathbf{k}^{\prime})L^{-2}(\omega^{\prime},\mathbf{k}^{\prime}) \,\big)\nn\\
        &-L^{-1}(\omega,\mathbf{k})\big(\,-(\omega-\omega^{\prime})\omega^{\prime}\delta_{ij} L^{-1}(\omega^{\prime},\mathbf{k}^{\prime})\big)\,\nn\\
        &-L^{-1}(\omega,\mathbf{k})\big(\,-v_i(k_j-k_j^{\prime})\omega^{\prime} L^{-1}(\omega^{\prime},\mathbf{k}^{\prime})\big)\,\nn\\
        &-L^{-1}(\omega,\mathbf{k})\big(\,\mathbf{v}\cdot(\mathbf{k}-\mathbf{k}^{\prime})\omega^{\prime}\delta_{ij} L^{-1}(\omega^{\prime},\mathbf{k}^{\prime})\big)\,\nn\\
        &+L^{-1}(\omega,\mathbf{k})\big(\,-v_iv_j\omega^{\prime}(\omega-\omega^{\prime})L^{-1}(\omega^{\prime},\mathbf{k}^{\prime}) \,\big)\nn\\
        &+\frac{p^0}{\chi T_0\tau_R}L^{-1}(\omega,\mathbf{k})\big(\,iv_i(\omega-\omega^{\prime})L^{-1}(\omega^{\prime},\mathbf{k}^{\prime})G_R^{0,j}(\omega^{\prime},\mathbf{k}^{\prime}) \,\big)\nn\\
        &+\frac{p^0}{T_0}L^{-1}(\omega,\mathbf{k})\big(\,-v_iv_j\omega^{\prime}(\omega-\omega^{\prime})L^{-1}(\omega^{\prime},\mathbf{k}^{\prime}) \,\big)\nn\\
        &+[i\leftrightarrow j, \omega-\omega^{\prime}\leftrightarrow \omega^{\prime}, \mathbf{k}-\mathbf{k}^{\prime}\leftrightarrow\mathbf{k}^{\prime}]\Big) .\label{GJ033}
\end{align}

\section{Basis of angular integrals}
\label{integrals}

We consider integrals of the form
\[
I_{mnl} = \int \frac{d\Omega}{2\pi}
\frac{(\mathbf{v}\cdot\mathbf{k})^m(\mathbf{v}\cdot\mathbf{k}')^n}
{(a + i\,\mathbf{v}\cdot\mathbf{k})(b + i\,\mathbf{v}\cdot\mathbf{k}')^l},
\]
where \(m,n\ge0\), \(m+n\le3\), \(a\equiv \frac{1}{\tau_R}-i\omega\), \(b\equiv \frac{1}{\tau_R}-i\omega'\).
All integrals required in the main text reduce algebraically to this basis.

Let
\[
A \equiv a + i\,\mathbf{v}\cdot\mathbf{k},\qquad
B \equiv b + i\,\mathbf{v}\cdot\mathbf{k}'.
\]
Using
\[
\mathbf{v}\cdot\mathbf{k} = -i(A-a),\qquad
\mathbf{v}\cdot\mathbf{k}' = -i(B-b).
\]

The numerator can be expanded as
\[
(\mathbf{v}\cdot\mathbf{k})^m(\mathbf{v}\cdot\mathbf{k}')^n
= (-i)^{m+n} (A-a)^m (B-b)^n.
\]
Using the binomial theorem:
\[
(A-a)^m = \sum_{p=0}^{m} \binom{m}{p} A^{p} (-a)^{m-p},
\]
\[
(B-b)^n = \sum_{q=0}^{n} \binom{n}{q} B^{q} (-b)^{n-q}.
\]
Thus the integral becomes
\[
I_{mnl} = (-i)^{m+n}
\sum_{p=0}^{m}\sum_{q=0}^{n}
\binom{m}{p}\binom{n}{q}
(-a)^{m-p} (-b)^{n-q}
\int \frac{d\Omega}{2\pi} A^{p-1} B^{q-l}.
\]
Thus the angular dependence reduces to master integrals of the form
\[
M_{pql} \equiv \int \frac{d\Omega}{2\pi} A^{p-1} B^{q-l}.
\]

\section{Explicit evaluation of \texorpdfstring{$I_{001}$}{I001}}
\label{app:I001}

The mixed scalar master integral is
\[
I_{001} = \int \frac{d\Omega}{2\pi} \frac{1}{(a + i\,\mathbf{v}\cdot\mathbf{k})(b + i\,\mathbf{v}\cdot\mathbf{k}')},
\]
where $|\mathbf v|=1$ and the integration is over its direction.  $I_{002}$ follows by differentiation,
\begin{align}
    I_{002}=-\frac{\partial I_{001}}{\partial b}=-\frac{\partial I_{001}}{\partial (-i\omega^\prime)}=-i\frac{\partial I_{001}}{\partial \omega^\prime}.
\end{align}

Use the standard Feynman parametrization
\[
\frac{1}{XY} = \int_0^1 \frac{dt}{\bigl[tX + (1-t)Y\bigr]^2}.
\]
Set
\[
X = a + i\,\mathbf{v}\cdot\mathbf{k},\qquad Y = b + i\,\mathbf{v}\cdot\mathbf{k}'.
\]
Then
\[
tX + (1-t)Y = t(a + i\,\mathbf{v}\cdot\mathbf{k}) + (1-t)(b + i\,\mathbf{v}\cdot\mathbf{k}')
= \alpha(t) + i\,\mathbf{v}\cdot\mathbf{w}(t),
\]
where
\[
\alpha(t) \equiv t a + (1-t)b,\qquad \mathbf{w}(t) \equiv t\mathbf{k} + (1-t)\mathbf{k}'.
\]
Hence,
\[
I_{001} = \int_0^1 dt \int \frac{d\Omega}{2\pi} \frac{1}{\bigl(\alpha(t) + i\,\mathbf{v}\cdot\mathbf{w}(t)\bigr)^2}.
\]

For fixed $t$, define $\beta = |\mathbf{w}(t)|$. We need
\[
J(\alpha,\beta) \equiv \int \frac{d\Omega}{2\pi} \frac{1}{(\alpha + i\,\mathbf{v}\cdot\mathbf{w})^2}.
\]
Let $u = \frac{\mathbf{v}\cdot\mathbf{w}}{\beta}$, so $u \in [-1,1]$ and $\frac{d\Omega}{2\pi} =  du$. Then
\[
J =  \int_{-1}^1 \frac{du}{(\alpha + i\beta u)^2}
=  \left[ \frac{1}{i\beta} \frac{-1}{\alpha + i\beta u} \right]_{-1}^1
= \frac{1}{ i\beta} \left( \frac{1}{\alpha - i\beta} - \frac{1}{\alpha + i\beta} \right)
= \frac{2}{\alpha^2 + \beta^2}.
\]
Thus
\[
I_{001} = 2\int_0^1 \frac{dt}{\alpha(t)^2 + \beta(t)^2}.
\]

Let
\[
D(t) \equiv \alpha(t)^2 + \beta(t)^2.
\]
Since $\alpha(t)$ is linear in $t$ and $\beta^2(t) = |\mathbf{w}(t)|^2$ is quadratic in $t$, $D(t)$ is a quadratic polynomial:
\[
D(t) = D_2 t^2 + D_1 t + D_0.
\]
with
\[
\begin{aligned}
D_2 &= (a-b)^2 + |\mathbf{k}-\mathbf{k}'|^2,\\
D_1 &= 2b(a-b) + 2(\mathbf{k}-\mathbf{k}')\cdot \mathbf{k}',\\
D_0 &= b^2 + |\mathbf{k}'|^2.
\end{aligned}
\]

The remaining  integral is elementary and yields
\[
I_{001} =  -\frac{4\left(
\arctan\!\left(\frac{D_1}{\sqrt{-D_1^2+4D_0D_2}}\right)
-
\arctan\!\left(\frac{D_1+2D_2}{\sqrt{-D_1^2+4D_0D_2}}\right)
\right)}
{\sqrt{-D_1^2+4D_0D_2}}.
\]

\section{The unified expansion formula \texorpdfstring{$M_{pql}$}{Mpql}}
\label{app:M}

The remaining angular integrals are generated from
\[
M_{pql} = \int \frac{d\Omega}{2\pi} A^{p-1} B^{q-l}.
\]
Thus we have
\begin{align}
    M_{00l}=I_{00l}, \quad M_{012}=I_{001},
\end{align}
and
\begin{align}
    M_{111}=M_{122}=2.
\end{align}

\subsection{Calculation of \texorpdfstring{$M_{011}$, $M_{022}$, $M_{101}$ and $M_{112}$}{M011, M022, M101 and M112}}

We calculate $M_{011}$, $M_{022}$, $M_{101}$ and $M_{112}$,
\begin{align}
M_{011}&=M_{022}= \int \frac{d\Omega}{2\pi} \frac{1}{a + i\,\mathbf{v}\cdot\mathbf{k}}=\frac{1}{i\,|\mathbf{k}|} \ln\!\left( \frac{a + i|\mathbf{k}|}{a - i|\mathbf{k}|} \right), \\
M_{101}=M_{112}&=M_{011}/\{a\rightarrow b, \, \mathbf{k}\rightarrow \mathbf{k}^\prime\} .
\end{align}

\subsection{Calculation of \texorpdfstring{$M_{102}$}{M102}}

For the double pole in the second denominator,
\begin{align}
M_{102}= \int \frac{d\Omega}{2\pi} \frac{1}{(b + i\,\mathbf{v}\cdot\mathbf{k}^\prime)^2}= \frac{2}{b^2 + |\mathbf{k}'|^2}.
\label{M102}
\end{align}

\subsection{Calculation of \texorpdfstring{$M_{010}$, $M_{021}$ and $M_{032}$}{M010, M021 and M032}}

Finally, the three master integrals needed with one power of $B$ in the numerator coincide:
\[
M_{032}=M_{021}=M_{010} \equiv \int \frac{d\Omega}{2\pi} \frac{b + i\,\mathbf{v}\cdot\mathbf{k}'}{a + i\,\mathbf{v}\cdot\mathbf{k}}.
\]

Decompose $\mathbf{k}'$ into components parallel and perpendicular to $\mathbf{k}$:
\begin{equation}
\mathbf{k}' = \left( \frac{\mathbf{k}\cdot\mathbf{k}'}{k^2} \right) \mathbf{k} + \mathbf{k}'_\perp \quad (\text{where } \mathbf{k}'_\perp \cdot \mathbf{k} = 0)
\end{equation}

The component perpendicular to $\mathbf k$ integrates to zero by azimuthal symmetry:
\begin{equation}
\int \frac{d\Omega}{2\pi} \frac{i\,\mathbf{v}\cdot\mathbf{k}'_\perp}{a + i\,\mathbf{v}\cdot\mathbf{k}} = 0.
\end{equation}

Using $i\mathbf v\cdot\mathbf k=(a+i\mathbf v\cdot\mathbf k)-a$ gives
\begin{align}
 & \int \frac{d\Omega}{2\pi} \frac{b + i\,\mathbf{v}\cdot\mathbf{k}'}{a + i\,\mathbf{v}\cdot\mathbf{k}} \nonumber \\
&= \left( \frac{\mathbf{k}\cdot\mathbf{k}'}{k^2} \right) \int \frac{d\Omega}{2\pi} (1) + \left[ b - a \left( \frac{\mathbf{k}\cdot\mathbf{k}'}{k^2} \right) \right] \int \frac{d\Omega}{2\pi} \frac{1}{a + i\,\mathbf{v}\cdot\mathbf{k}}.
\end{align}

Hence,
\begin{equation}
M_{032}=M_{021}=M_{010} =\int \frac{d\Omega}{2\pi} \frac{b + i\,\mathbf{v}\cdot\mathbf{k}'}{a + i\,\mathbf{v}\cdot\mathbf{k}} = 2 \left( \frac{\mathbf{k}\cdot\mathbf{k}'}{k^2} \right) + \frac{1}{i k} \left[ b - a \left( \frac{\mathbf{k}\cdot\mathbf{k}'}{k^2} \right) \right] \ln\left( \frac{a + i k}{a - i k} \right).
\end{equation}
Although the last representation contains terms proportional to $(\mathbf k\cdot\mathbf k')/k^2$, the apparent small-$k$ singularity cancels between the two terms.  Expanding the logarithm before taking the limit gives $M_{010}=M_{021}=M_{032}\to 2b/a$ as $k\to0$, providing a useful regularity check.

\bibliographystyle{apsrev4-2}
\bibliography{ref}

\end{document}